\documentclass[12pt,a4paper]{cimento}

\usepackage{amsmath}
\usepackage{gensymb}
\usepackage{enumitem}
\usepackage{graphicx}  

\title{Detection of the liquid-liquid transition in the deeply-cooled water\ETC\\ confined in MCM-41 with elastic neutron scattering technique}
\author{Zhe~Wang\from{ins:x}\ETC,
Kanae~Ito\from{ins:x}
   \atque
Sow-Hsin~Chen\from{ins:x}\thanks{sowhsin@mit.edu}}
\instlist{\inst{ins:x}Department of Nuclear Science and Engineering, Massachusetts Institute of Technology - Cambridge, MA, USA}

\PACSes{\PACSit{25.40.Dn}{Scattering}
\PACSit{64.70.Ja}{Liquids}}

\begin{document}

\maketitle

\begin{abstract}
In this paper we present a review on our recent experimental investigations into the phase behavior of the deeply-cooled water confined in a nanoporous silica material, MCM-41, with elastic neutron scattering technique. Under such strong confinement, the homogeneous nucleation process of water is avoided, which allows the confined water to keep as liquid state at temperatures and pressures that are inaccessible to the bulk water. By measuring the average density of the confined heavy water, we observe a likely first-order low-density liquid (LDL) to high-density liquid (HDL) transition in the deeply-cooled region of the confined heavy water. The phase separation starts from 1.12$\pm$0.17 kbar and 215$\pm$1 K and extends to higher pressures and lower temperatures in the phase diagram. This starting point could be the liquid-liquid critical point of the confined water. The locus of the Widom line is also estimated. The observation of the liquid-liquid transition in the confined water has potential to explain the mysterious behaviors of water at low temperatures. In addition, it may also have impacts on other disciplines, because the confined water system represents many biological and geological systems where water resides in nanoscopic pores or in the vicinity of hydrophilic or hydrophobic surfaces.
\end{abstract}

\section{Introduction}
Water is a continuing source of fascination to scientists not only due to its tremendous political, cultural and historical significance, but also because of its anomalous physical behaviors. It is well-known that water has a density maximum at 4\degree{}C under ambient pressure. In fact, when cooling down, the thermodynamic response functions and transport coefficients of water also exhibit counterintuitive behaviors \cite{ref1, ref2, ref3, ref4}. Moreover, the glassy water, also called amorphous ice, exhibits polyamorphism. Experiments show that two kinds of amorphous ice, the low-density amorphous ice (LDA) and the high-density amorphous ice (HDA), exist at very low temperatures \cite{ref5, ref6, ref7, ref8}. These two phases can transform to each other through a first-order transition \cite{ref7, ref8}.

\begin{figure}
	\centering
\includegraphics{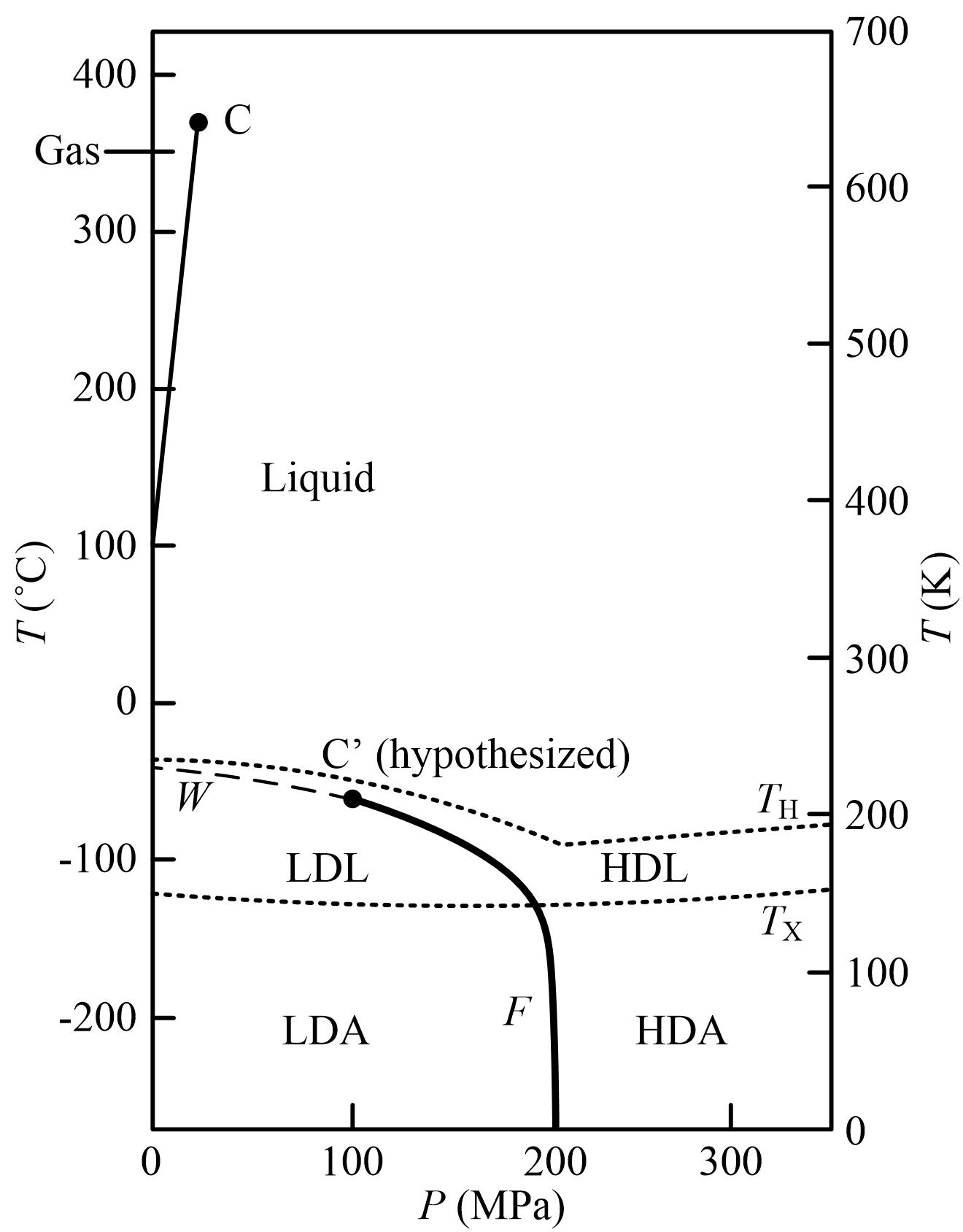}
\caption{The phase diagram of the LLCP scenario. C and C' denote the known liquid-gas critical point and the hypothesized liquid-liquid critical point respectively. $F$ denotes the line of first-order phase transitions that emanates from C' and separates the high-density and low-density phases that occur for temperatures below $T_\mathrm{c'}$. $W$ denotes the Widom line. $T_\mathrm{H}$ denotes the homogeneous nucleation temperature line. $T_\mathrm{X}$ denotes the crystallization temperatures of amorphous ice. The region between $T_\mathrm{H}$ and $T_\mathrm{X}$ is called ``no man's land'', because in this region bulk water cannot exist as liquid state. From O. Mishima, and H. E. Stanley, Nature (London) \textbf{396}, 329 (1998).}
\end{figure}

To account for these mysterious phenomena, Eugene Stanley and his collaborators proposed a theoretical picture call liquid-liquid critical point (LLCP) scenario \cite{ref9}. It hypothesizes that water has two liquid phases at low temperatures: low-density liquid (LDL) phase and high-density liquid (HDL) phase. Figure 1 shows the schematic phase diagram of the LLCP scenario \cite{ref4}. It can be found that, in this scenario, the LDL and HDL phases are thermodynamic extensions of the LDA and HDA phases into the liquid state, respectively. The transition between LDL and HDL is a first-order phase transition. This liquid-liquid phase transition ends at a critical point, which is called liquid-liquid critical point (LLCP). The Widom line is the extension of the liquid-liquid transition line into the one-phase region. It can be defined as the locus of the maximum of correlation length \cite{ref10} or the locus of the maximum of isobaric heat capacity \cite{ref11}.

Testing the existence of the liquid-liquid transition is crucial for understanding the low-temperature behaviors of water. Unfortunately, such experiments are practically difficult, since these two liquid phases are supposed to exist at temperatures lower than the homogeneous nucleation temperature $T_\mathrm{H}$ (232 K at ambient pressure) where bulk water cannot stay as liquid (for this reason, this low-temperature region is called ``no man's land''). To overcome this barrier, a hydrophilic nanoporous silica material, MCM-41, is used to confine the water. Such strong confinement can suppress the homogeneous nucleation process, so that it can keep the confined water in liquid state at temperatures even below $T_\mathrm{H}$. Subsequently, the confined water system provides an opportunity to investigate the behaviors of the liquid water in the deeply-cooled region (the word ``deeply-cooled'' describes the region at temperatures below bulk $T_\mathrm{H}$). Notice that, the confined water can suffer from constraints (geometrical and chemical) induced by the confinement \cite{ref12, ref13, ref14}. Therefore, to what extent the confined water is similar to the bulk water is still in debate. However, such a confined water system is of fundamental importance in practice and fascinates scientists from different disciplines. For example, it represents many biological and geological systems where water resides in nanoscopic pores or in the vicinity of hydrophilic or hydrophobic surfaces.

The aim of this paper is to present our effort in the detection of the liquid-liquid transition in the water confined in MCM-41 with elastic neutron scattering technique. We will first introduce two prerequisites for further discussions, namely, the properties of the MCM-41 sample and the model for extracting the average density of the confined water from elastic neutron scattering measurement. Then we will discuss the establishment of the phase diagram of the liquid-liquid transition in the confined water system. The content of this paper is organized as follows.

\begin{enumerate}
	\item[] Summary
	\item Introduction
	\item Experimental Methods
	\begin{enumerate}[label*=\arabic*.]
		\item MCM-41
		\item Model Description
	\end{enumerate}
	\item Results and Discussions
	\begin{enumerate}[label*=\arabic*.]
		\item Phase Diagram
		\item Partially-Hydrated Sample		
	\end{enumerate}
	\item Concluding Remarks
\end{enumerate}

\section{Experimental Methods}
\subsection{MCM-41}
MCM-41 is a mesoporous silica material. It is made by calcining self-assembled micellar templated silica matrices, which are composed of grains of micrometer size. In each grain, parallel and uniform-sized cylindrical pores are arranged in a well-ordered two-dimensional hexagonal lattice. MCM-41 has hydrophilic surface and large pore volume to confine sufficient amount of water, and also small enough pore size to inhibit the homogeneous nucleation process of water. From a series of differential scanning calorimeter (DSC) measurements, we confirmed that when the nominal pore size is smaller than $\sim$17 \AA, the ice nucleation can be bypassed and the confined water can be supercooled at least down to $\sim$130 K without freezing \cite{ref15}.

Figure 2 shows the two-dimensional hexagonal geometry of pores in MCM-41. Distances ``\emph{a}'' and ``\emph{d}'' represent the inter-pore and inter-plane distances respectively.

\begin{figure}
	\centering
\includegraphics{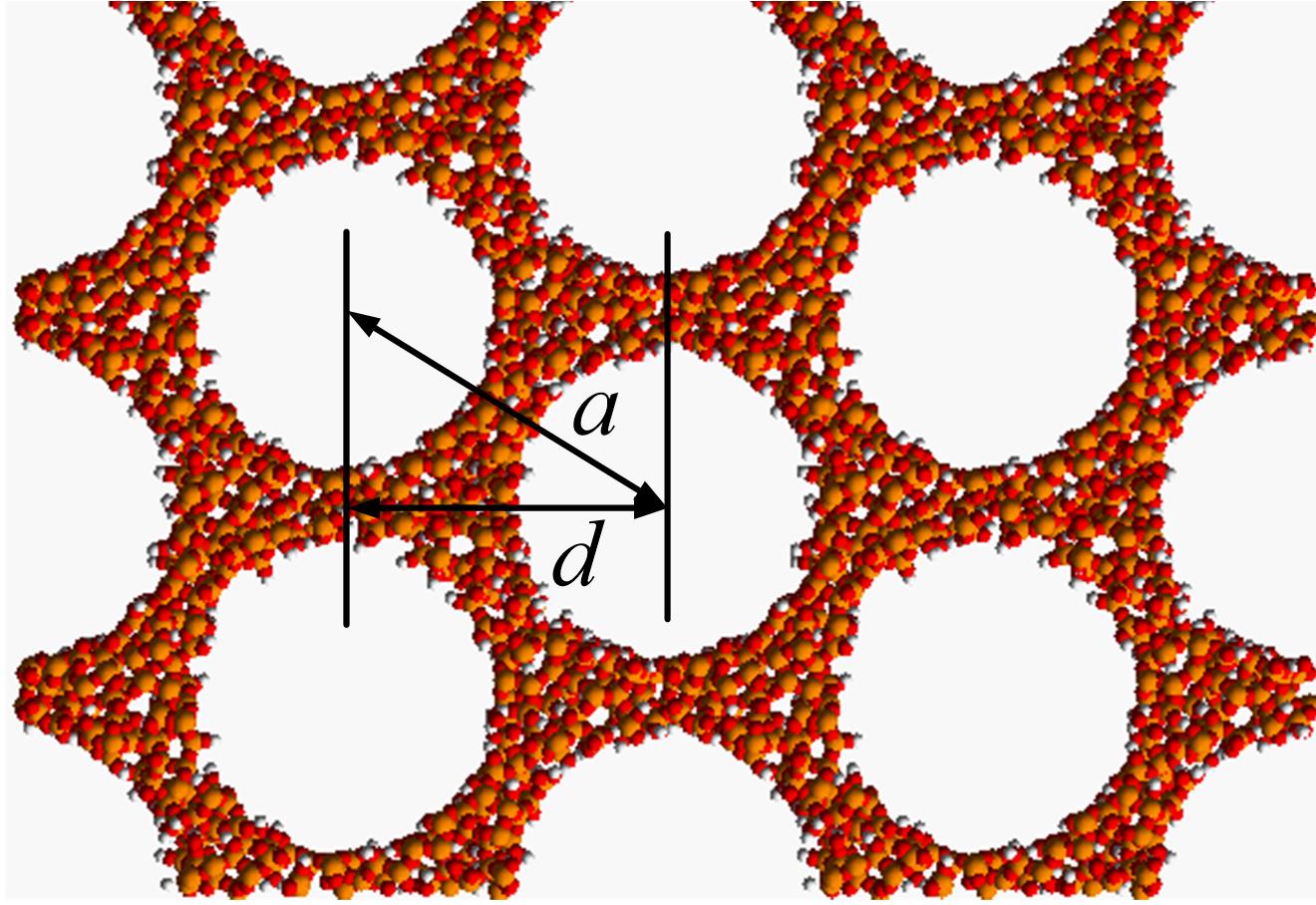}
\caption{Two-dimensional hexagonal pore structure of MCM-41 used in this project. The center-to-center distance between two adjacent pores is marked as ``$a$'', and the distance between two adjacent rows is marked as ``$d$''. Notice that $d = \sqrt{3}a/2$, which equals to 29 \AA.}
\end{figure}

In our studies, MCM-41 with nominal pore size of 15$\sim$16~\AA~was used as the confining matrix. The pore size distribution is estimated by Barret-Joyner-Halenda (BJH) method using nitrogen sorption isotherms \cite{ref16, ref17}. Water can be introduced into the pores via vapor condensation easily. The full hydration level by weight $h$ (g water/g dry MCM-41) is about 0.45 g/g for H$_2$O-hydrated sample, and 0.50 g/g for D$_2$O-hydrated sample. The adsorption isotherm data presented in Fig. 3 justifies these designations.

\begin{figure}
	\centering
\includegraphics{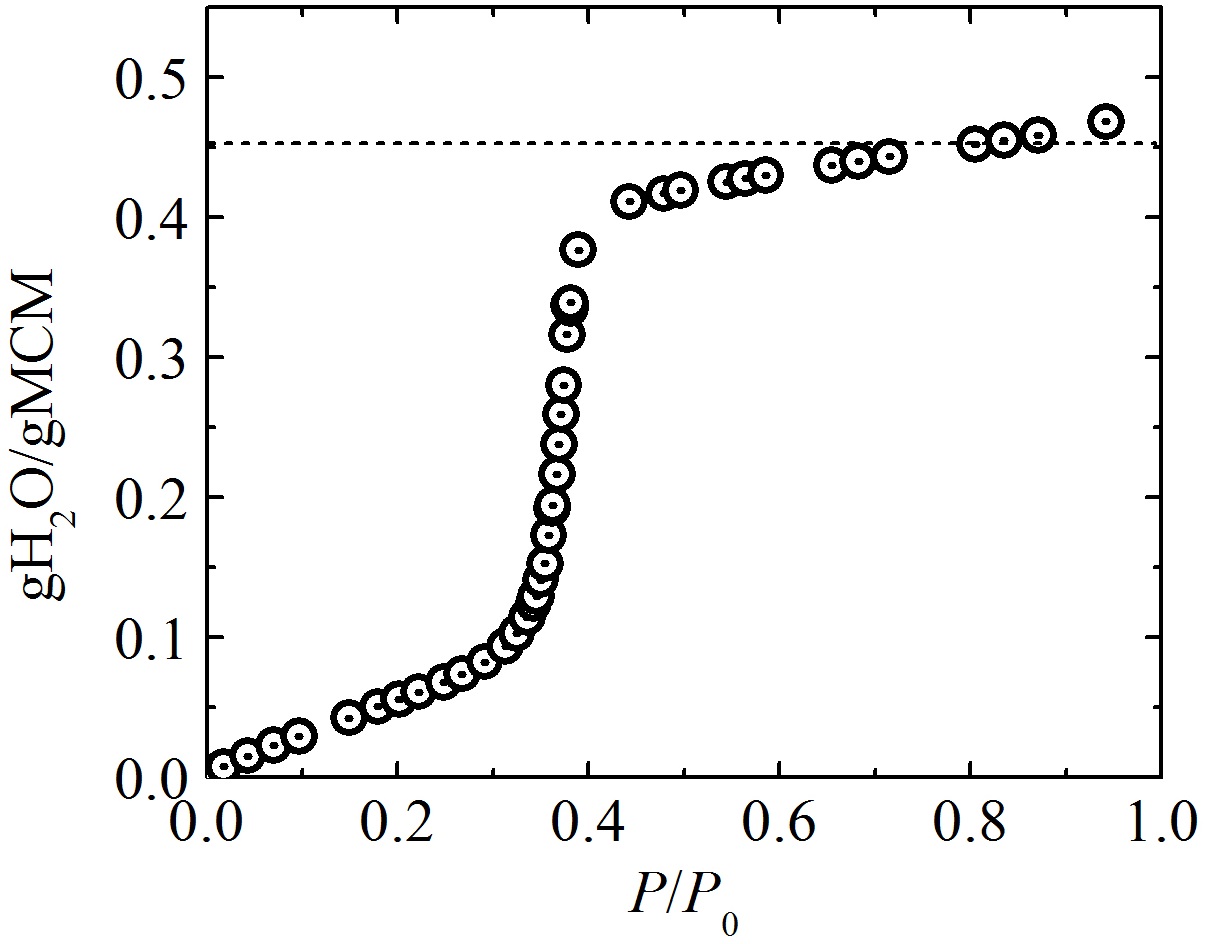}
\caption{Isothermal adsorption of water vapor (H$_2$O) onto MCM-41 at room temperature. $P_0$ denotes the ambient pressure. The horizontal dotted line marks the full hydration level of the sample.}
\end{figure}

\subsection{Model Description}
The order parameter of the hypothetical liquid-liquid transition is density. Therefore, it is important to know how to measure the average density of the water confined in MCM-41. In 2007, Liu \emph{et al}. developed a method for the measurement of the average density of the confined heavy water with elastic neutron scattering technique \cite{ref18}. In this section we will introduce this method in detail.

As shown in Fig. 2, the two-dimensional structure of MCM-41 has a hexagonal order. This order will produce a Bragg peak in the neutron diffraction spectrum of the confined water system. The center of the Bragg peak locates at $Q = 2\pi/d = 0.21 $\AA$^{-1}$. A typical elastic neutron scattering spectrum of the confined heavy water system is shown in Fig. 4.

\begin{figure}
	\centering
\includegraphics{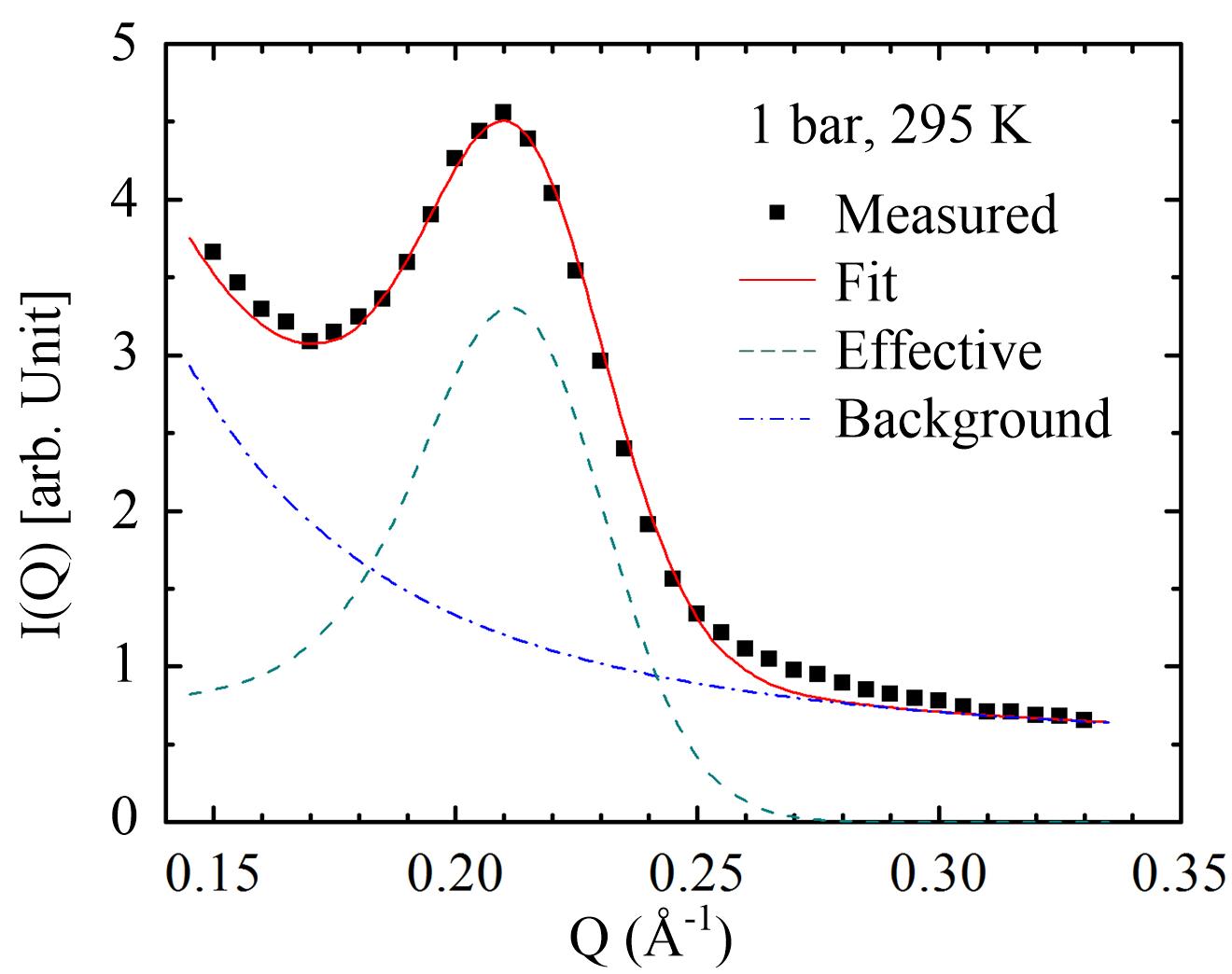}
\caption{Typical elastic neutron scattering spectrum of the confined heavy water system (denoted by black solid squares) at 1 bar and 295 K. The red curve is the fitting curve with Eq. 2. The green dashed line represents the ``effective'' part of the spectrum, \emph{i.e.}, the first term in the right-hand-side part of Eq. 2. The dash-dot line represents the background, \emph{i.e.}, the terms $BQ^{-\beta}+C$ in Eq. 2.}
\end{figure}

The measured neutron diffraction spectrum of the confined heavy water system consists of three parts: (i) the low-$Q$ scattering of the fractal packing of the MCM-41 grains and the background signal due to the low-$Q$ nature of the instrument; (ii) the $Q$-independent incoherent background and (iii) a Bragg peak at $2\pi/d$ due to the two-dimensional hexagonal lattice of the MCM-41. The first part is represented by $BQ^{-\beta}$. The second part is represented by a constant $C$. The third part is given by $nV_p^2(\Delta\rho^\mathrm{sld})^2\bar{P}(Q)S(Q)$, where $n$ is the number of scattering units (water cylinders) per unit volume, $V_p$ is the volume of the scattering unit, $\Delta\rho^\mathrm{sld}=\rho_\mathrm{D_2O}^\mathrm{sld}-\rho_\mathrm{MCM}^\mathrm{sld}$ is the difference of the scattering length density (sld) between the scattering unit (D$_2$O cylinder) and the environment (MCM-41). $\bar{P}(Q)$ is the form factor of the scattering unit. $S(Q)$ is the inter-cylinder structure factor of the two-dimensional hexagonal lattice. The sld of the scattering unit (D$_2$O cylinder) $\rho_\mathrm{D_2O}^\mathrm{sld}$ is proportional to the average density of the confined D$_2$O~$\rho_\mathrm{D_2O}^\mathrm{m}:\rho_\mathrm{D_2O}^\mathrm{sld}=\alpha\rho_\mathrm{D_2O}^\mathrm{m}$ , where $\alpha=N_A\sum{b_i/M}$, $N_A$ is Avogadro's number, $M$ is the molecular weight of D$_2$O and $b_i$ is the coherent scattering length of the $i$th atom in the scattering unit. The form factor $\bar{P}(Q)$ for a long cylinder ($QL > 2\pi$) is given by $(\pi/QL)(2J_1(QR)/QR)^2$, where $R$ is the radius of the cylinder, $L$ is the length of the cylinder and $J_1(x)$ is the first-order Bessel function of the first kind. The inter-cylinder structure factor of the two-dimensional hexagonal lattice $S(Q)$ could be modeled by a Lorentzian function. Therefore the $Q$ scan intensity distribution of the system can be modeled as:
\begin{multline}
	\label{eq:eq1}
	I(Q) = nV_p^2(\alpha\rho_\mathrm{D_2O}^\mathrm{m}-\rho_\mathrm{MCM}^\mathrm{sld})^2\frac{\pi}{QL}\left[\frac{2J_1(QR)}{QR}\right]^2\left[\frac{\frac{1}{2}\Gamma}{(Q-\frac{2\pi}{d})^2+(\frac{1}{2}\Gamma)^2}\right]\\+BQ^{-\beta}+C.
\end{multline}
Eq. 1 can be rewritten as:
\begin{equation}
	\label{eq:eq2}
	I(Q) = A\frac{J_1(QR)^2}{Q^3R^2}\left[\frac{\frac{1}{2}\Gamma}{(Q-\frac{2\pi}{d})^2+(\frac{1}{2}\Gamma)^2}\right]+BQ^{-\beta}+C.
\end{equation}

where $A$ is expressed as:
\begin{equation}
	\label{eq:eq3}
	\begin{split}
	A & =nV_p^2\alpha^2\left(\rho_\mathrm{D_2O}^\mathrm{m}-\frac{\rho_\mathrm{MCM}^\mathrm{sld}}{\alpha}\right)^2\frac{4\pi}{L}\\
	  & =A_1\left(\rho_\mathrm{D_2O}^\mathrm{m}-\frac{\rho_\mathrm{MCM}^\mathrm{sld}}{\alpha}\right)^2.
	\end{split}
\end{equation}
Notice that the average density of the confined D$_2$O $\rho_\mathrm{D_2O}^\mathrm{m}$ is contained in $A$. The value of $A$ can be obtained by convoluting Eq. 2 with the instrument resolution and fitting it to the measured $Q$ scan data. Figure 4 shows the fitting curve using this model. In order to determine $\rho_\mathrm{D_2O}^\mathrm{m}$ from $A$, one needs to know the values of $A_1$, $\rho_\mathrm{MCM}^\mathrm{sld}$ and $\alpha$. $\rho_\mathrm{MCM}^\mathrm{sld}$ and $\alpha$ can be obtained by contrast variation \cite{ref18}. $A_1$ can be found by comparing the experimental data with a previous result on the absolute value of the density of the confined water \cite{ref19}.

In Eq. 2, $R$, $B$, $C$ and $\beta$ are constants. The value of $d$, which reflects the structure of the MCM-41, depends on temperature and pressure very weakly. This is because (\emph{i}) the thermal expansion coefficient of the MCM-41 is only in the order of 10$^{-6}$ /K, which is smaller than that of the water by three orders; (\emph{ii}) as a solid, the MCM-41 exhibits very small compressibility. Considering that $d$ is almost a constant, one can find that at $Q = 0.21$ \AA$^{-1}$ (we denote this $Q$ value as $Q_\mathrm{B}$ in the following part since it is close to the position of the Bragg peak $2\pi/d$) the $Q$ scan intensity is expressed as:
\begin{equation}
	\label{eq:eq4}
	\begin{split}
	I(Q_\mathrm{B}) & = nV_p^2(\alpha\rho_\mathrm{D_2O}^\mathrm{m}-\rho_\mathrm{MCM}^\mathrm{sld})^2\frac{\pi}{Q_\mathrm{B}L}\left[\frac{2J_1(Q_\mathrm{B}R)}{Q_\mathrm{B}R}\right]^2\left[\frac{\frac{1}{2}\Gamma}{(Q_\mathrm{B}-\frac{2\pi}{d})^2+(\frac{1}{2}\Gamma)^2}\right]+BQ_\mathrm{B}^{-\beta}+C. \\
	& \approx nV_p^2 \frac{\pi}{Q_\mathrm{B}L}\left[\frac{2J_1(Q_\mathrm{B}R)}{Q_\mathrm{B}R}\right]^2\frac{2}{\Gamma}\alpha^2\left(\rho_\mathrm{D_2O}^\mathrm{m}-\frac{\rho_\mathrm{MCM}^\mathrm{sld}}{\alpha}\right)^2+BQ_\mathrm{B}^{-\beta}+C \\
	&=D\left(\rho_\mathrm{D_2O}^\mathrm{m}-\frac{\rho_\mathrm{MCM}^\mathrm{sld}}{\alpha}\right)^2+BQ_\mathrm{B}^{-\beta}+C
	\end{split}
\end{equation}
where
\begin{equation}
	D = nV_p^2 \frac{\pi}{Q_\mathrm{B}L}\left[\frac{2J_1(Q_\mathrm{B}R)}{Q_\mathrm{B}R}\right]^2\frac{2}{\Gamma}\alpha^2
\end{equation}
In Eq. 5, $\Gamma$ exhibits weak temperature and pressure dependences. Subsequently, the value of $D$ can be considered approximately as a constant in certain pressure and temperature ranges. Therefore, from the last step of Eq. 4 one can find that $I(Q_\mathrm{B})$ is a monotonic function of $\rho_\mathrm{D_2O}^\mathrm{m}$. Knowing this, one can use $I(Q_\mathrm{B})$ to monitor the change of $\rho_\mathrm{D_2O}^\mathrm{m}$.

Liu \emph{et al}. use this model to study the average density of the confined heavy water as a function of temperature at ambient pressure \cite{ref18}. They find a well-defined minimum point at 210 K, which is consistent with the computer simulation prediction \cite{ref20}. The density profile of the confined heavy water is shown in Fig. 5.

\begin{figure}
	\centering
\includegraphics{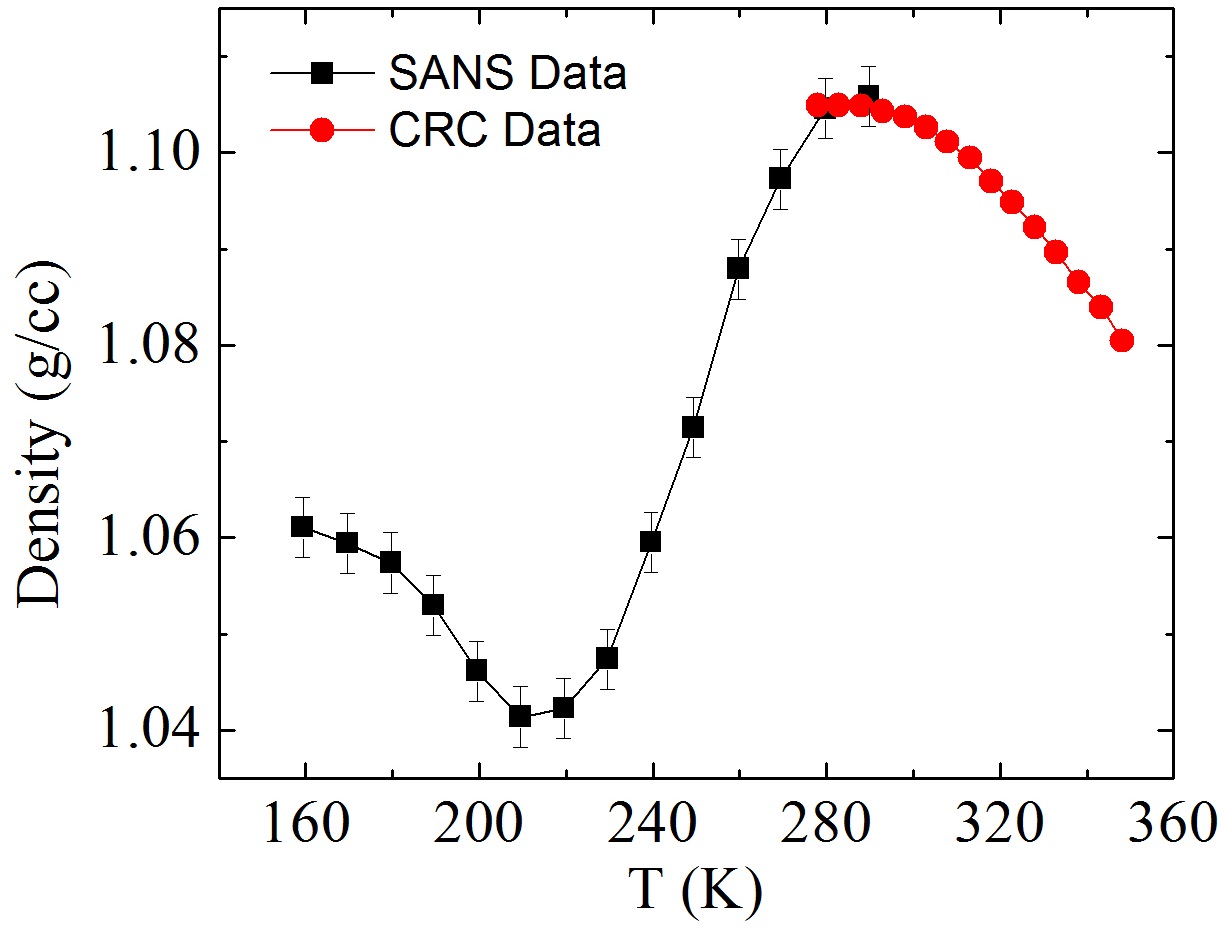}
\caption{Average density of the confined D$_2$O at ambient pressure obtained with small angle neutron scattering.  A smooth transition of D$_2$O density from the maximum value at 284 K to the minimum value at 210 K is clearly shown. The red circles are the density data for bulk D$_2$O taken from the CRC Handbook. From D. Liu, \emph{et al.}, Proc. Natl. Acad. Sci. U.S.A. \textbf{104}, 9570 (2007).}
\end{figure}

The existence of the density minimum in confined water is confirmed by light scattering \cite{ref21} and X-ray scattering \cite{ref15}. The observation of the density minimum is significant. Its occurrence would signal the reversal of the anomalies that set in near the density maximum; \emph{i.e.}, that mildly supercooled water is anomalous but that deeply supercooled water ``goes normal''. In addition, the existence of the density minimum in water is ascribed to the local tetrahedral structure, and is consistent with the existence of the liquid-liquid phase transition \cite{ref20}.

In the following part, we will introduce our work on the detection of the liquid-liquid transition in the confined heavy water with the method described in this section.

\section{Results and Discussions}
\subsection{Phase Diagram}
It is common that a first-order phase transition exhibits metastability. Therefore, one can test the existence of the hypothetical first-order liquid-liquid transition by detecting the hysteresis of the relevant order parameter, namely, the density of water. For example, one can measure the density of the confined water with warming scan and cooling scan at a specific pressure. If the experimental routes cross the phase boundary (solid arrow lines in Fig. 6), these two routes could give different density profiles as a function of temperature, due to the discontinuity at the phase boundary and the strong metastability of the liquid water in the coexisting region \cite{ref22} as the result of liquid-liquid transition \cite{ref23, ref24} and to the confinement \cite{ref25, ref26}. On the contrary, if the experimental routes are in the continuous region (dashed arrow lines in Fig. 6), no hysteresis should be observed.

\begin{figure}
	\centering
\includegraphics{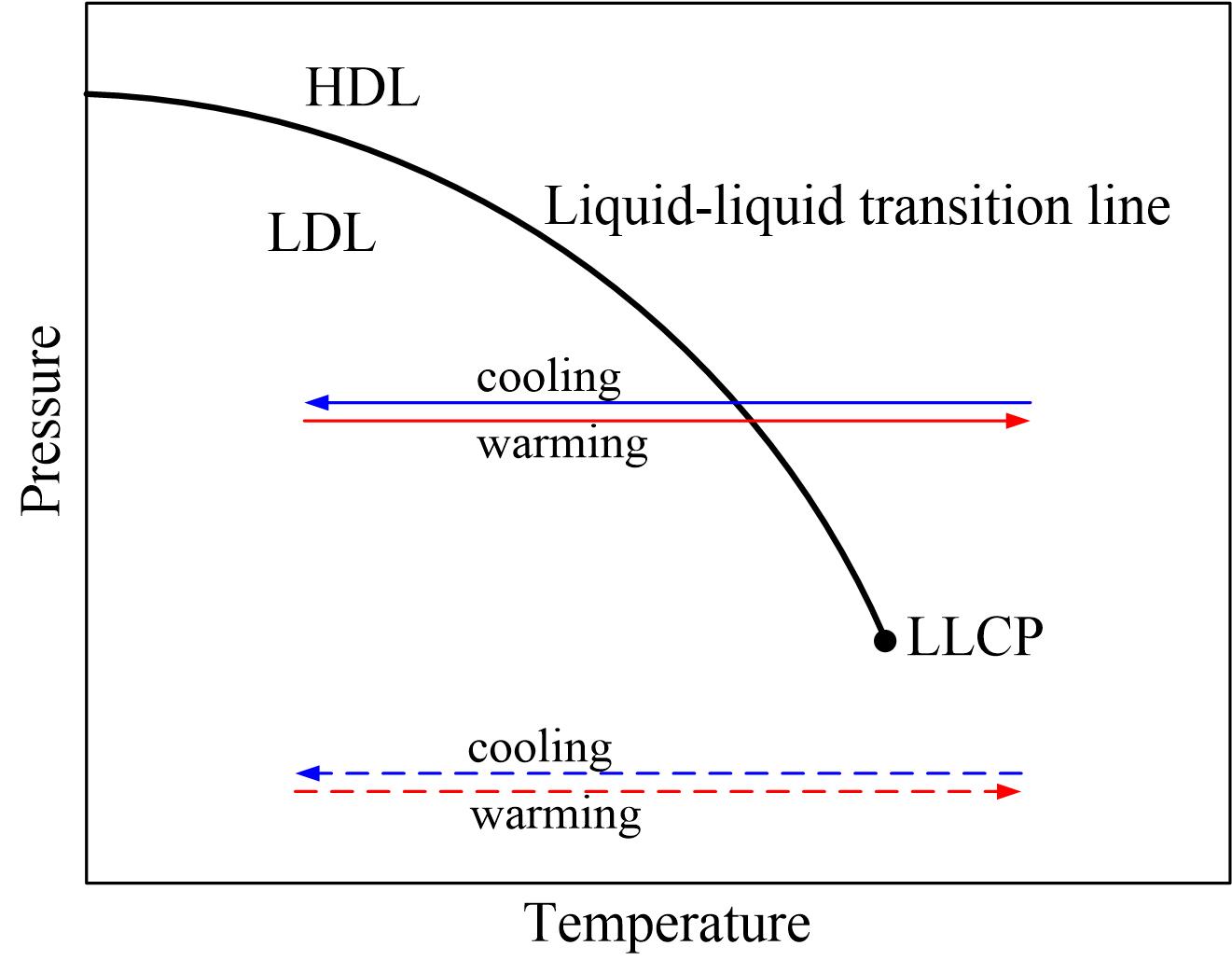}
\caption{Schematic phase diagram of the liquid-liquid transition and the experimental routes in the density hysteresis measurement. The black solid line denotes the liquid-liquid transition line. The warming and cooling scans at the higher pressure (solid arrow lines) cross the phase boundary, and should give a density hysteresis because of the long time required for the phase separation. However, at the lower pressure, there is no phase boundary, the warming and cooling scans should not give a density hysteresis.}
\end{figure}

In order to detect the density hysteresis in the confined water, we performed a series of neutron diffraction experiments to measure the average density of the confined heavy water with warming and cooling scans at different pressures. We used the following two protocols for the temperature scan:

(a) Continuous temperature scan. In this protocol, for each pressure, the sample was cooled from room temperature to 130 K at ambient pressure and then pressurized to the desired pressure. After two hours of waiting, the warming scan with 0.2 K/min was first performed from 130 to 300 K. When the warming scan was finished, we waited for another two hours and then performed the cooling scan with 0.2 K/min from 300 K to 130 K. During the temperature scan, the average density of the confined water was recorded for every minute.

(b) Discrete temperature scan. In this protocol, we did not change the temperature continuously as we did in protocol (a). On the contrary, for each scan, we only measure the density at several important temperatures. Before each density measurement, we wait for half an hour after the temperature reaches the desired value to guarantee the temperatures of the sample and the sample holder get equivalence.

The experiments employing protocol (a) were performed by Zhang \emph{et al}. first (in the pressure range from 1 bar to 3 kbar) \cite{ref27} and then followed by Wang \emph{et al}. (at pressures of 3.3 and 4 kbar) \cite{ref28, ref29}. The results are summarized in Fig. 7 \cite{ref30}.

\begin{figure}
	\centering
\includegraphics{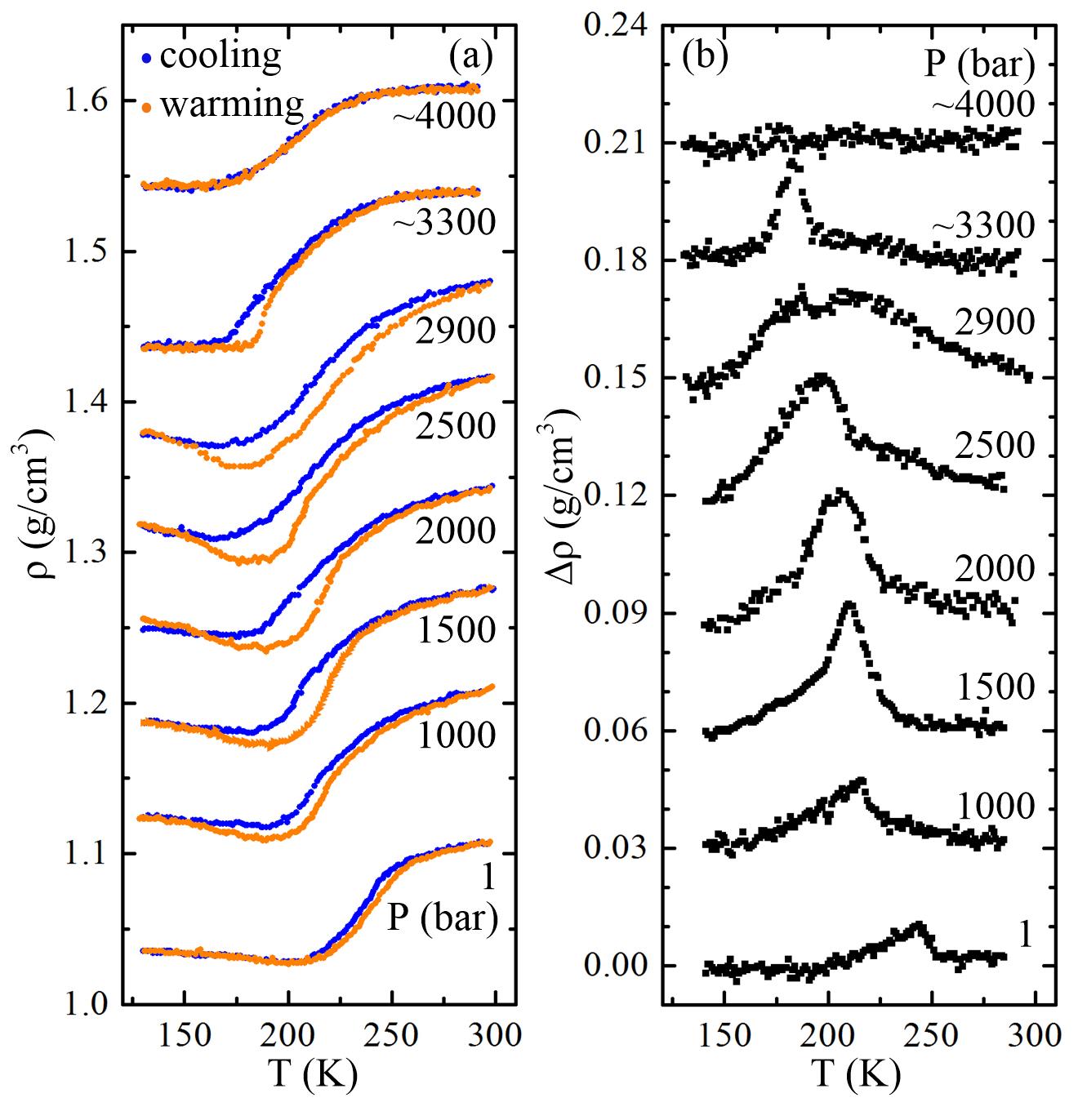}
\caption{The density measurement on the confined D$_2$O made by Zhang \emph{et al.} (1 bar to 2.9 kbar) \cite{ref27} and Wang \emph{et al.} (3.3 k and 4 kbar) \cite{ref28}. (a) The density profiles of confined D$_2$O with warming and cooling scans at different pressures. The data are shifted by 0.05 g/cm$^3$ between adjacent pressures for clarity. (b) The density differences between the cooling and warming scans at different pressures. The data are shifted by 0.03 g/cm$^3$ between adjacent pressures for clarity. From Z. Wang, \emph{et al.}, J. Phys. Chem. Lett. \textbf{6}, 2009 (2015).}
\end{figure}

The main results obtained from the experiments with the continuous temperature scans can be summarized as follows. (1) Density hysteresis phenomenon is observed at all the measured pressures below $\sim$3.5 kbar. (2) When the pressure is below $\sim$1.5 kbar, the hysteresis enhances as the pressure increases. The maximum density differences between the cooling and warming scans are 0.01 g/cm$^3$ at 1 bar, 0.017 g/cm$^3$ at 1 kbar, and 0.031 g/cm$^3$ at 1.5 kbar, respectively. (3) When the pressure is above $\sim$1.5 kbar, the amplitude of the hysteresis stabilizes at about 0.03 g/cm$^3$. (4) The temperature of the maximum density difference between the cooling and warming scans shifts to lower temperature as the pressure increases. The observation of the density hysteresis is significant, because it strongly suggests the existence of a first-order transition between a low-density phase and a high-density phase. Moreover, the feature that the hysteresis temperature decreases as the pressure increases qualitatively agrees with the $P$-$T$ dependence of the liquid-liquid transition line predicted by the computer simulation study \cite{ref9}. The phase diagram suggested by the experiments employing the continuous temperature scan is shown in Fig. 8.

\begin{figure}
	\centering
\includegraphics{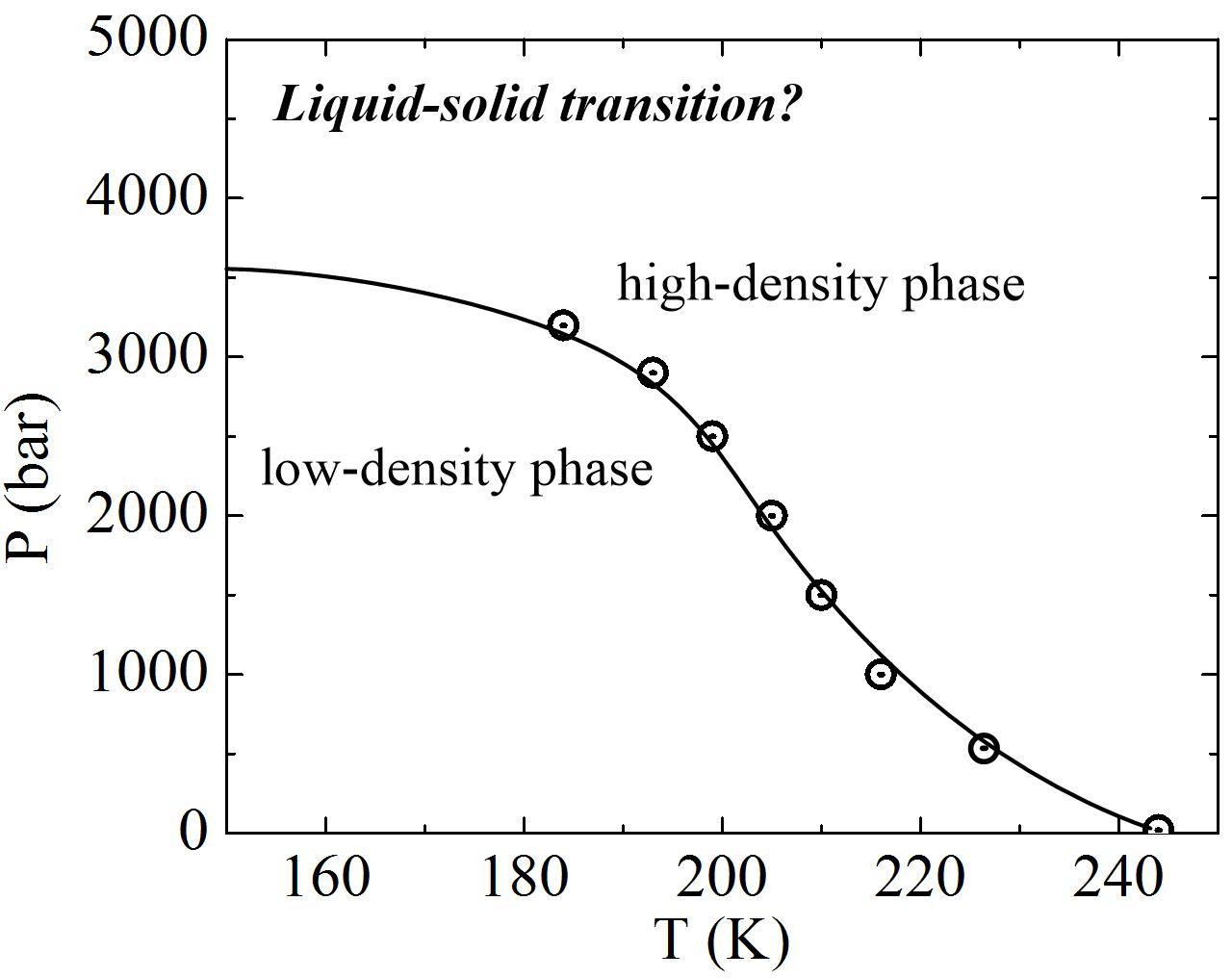}
\caption{Phase diagram suggested by the experiments employing the continuous temperature scan. The open circles denote the positions of the maximum density differences obtained by the continuous temperature scan. The solid line denotes the phase boundary between the low-density phase and the high-density phase. This picture suggests a liquid-solid transition rather than the liquid-liquid transition. However, the method for obtaining this diagram has a potential deficiency as explained in the text.}
\end{figure}

The observation of the density hysteresis was ascribed to the liquid-liquid transition by Zhang \emph{et al}.. However, this conclusion was soon challenged by Limmer and Chandler \cite{ref31}. With a computer simulation study employing mW model of water, these researchers attribute the density hysteresis phenomenon to a liquid-solid transition (LST) in the confined water (this result is in debate \cite{ref32, ref33}). An important difference between the LLCP scenario and LST scenario is that in the LLCP scenario there is a LLCP that terminates the liquid-liquid transition line at a positive pressure. In contrast, in the LST scenario there is no associated critical point and the liquid-solid transition line exists in all the positive pressures. From Figs. 7 and 8 one can find that, the density hysteresis appears even at 1 bar. Thus it seems that the LST scenario, rather than the LLCP scenario, provides a better explanation for the phase diagram shown in Fig. 8.

The continuous temperature scan protocol has a potential deficiency. The temperature changes continuously with a constant speed of 0.2 K/min. Though the speed is slow, it is possible that the heat transfer does not complete and the temperature sensor, which is on the aluminum holder of the sample, cannot accurately reflect the temperature of the confined water. Subsequently, there may be a temperature lag between the warming and cooling scans, and a hysteresis that is not due to the phase transition may appear. This problem can be solved by the discrete temperature scan protocol. As mentioned above, in this protocol, before recording the density, we wait for half an hour after the temperature reaches the desired value. Therefore, there is sufficient time for the sample to get a uniform temperature distribution and to reach temperature equivalence to the sample holder. The result of the density measurement with the discrete temperature scan protocol is shown in Fig. 9. It is found that the effective density hysteresis only appears when the pressure is higher than about 1000 bar. It takes place at the temperature that is very close to the one found in the experiments with continuous temperature scan. This result suggests a first-order transition between a low-density phase and a high-density phase, and is consistent with the LLCP picture, rather than the LST picture. The end point of the phase separation, which locates within $0.95<P<1.63$ kbar and $210<T<216$ K, is the LLCP of the confined D$_2$O according to the LLCP scenario. 

\begin{figure}
	\centering
\includegraphics{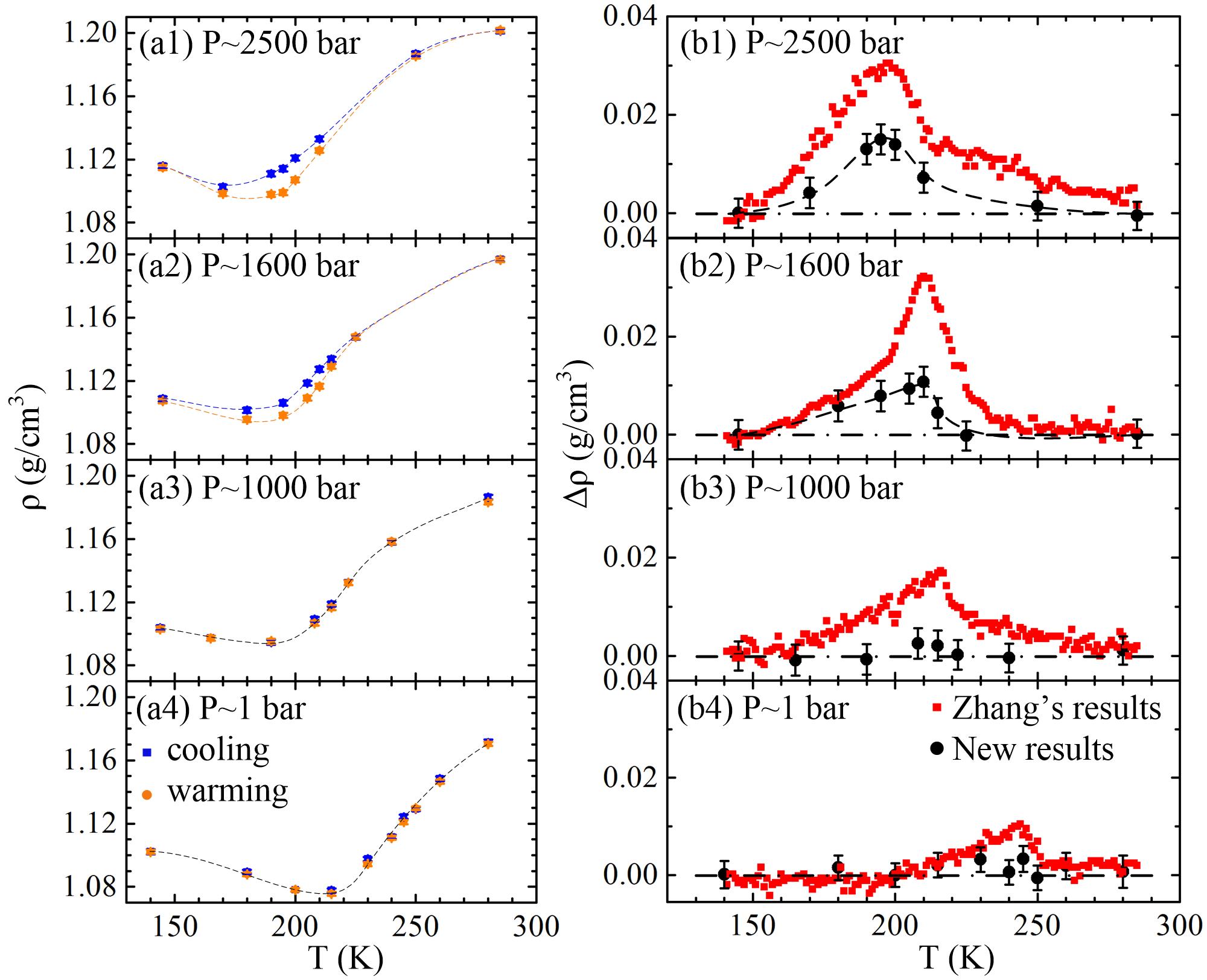}
\caption{The density measurement on the confined D$_2$O with the discrete temperature scan protocol. The left column shows the density profiles with warming and cooling scans at $P \sim$ 2500 bar (a1), 1600 bar (a2), 1000 bar (a3) and 1 bar (a4). The right column shows the density differences between the cooling and warming scans (denoted by black circles) at $P \sim$ 2500 bar (b1), 1600 bar (b2), 1000 bar (b3) and 1 bar (b4). We also plot the results of the density differences from the continuous temperature scans \cite{ref27} (denoted by red circles) for comparison. The dashed lines are drawn to guide eyes. From Z. Wang, \emph{et al.}, J. Phys. Chem. Lett. \textbf{6}, 2009 (2015).}
\end{figure}

For a more accurate position of the LLCP, we performed discrete warming and cooling scans at about 1.3 kbar. The result shows an effective density hysteresis with the amplitude of $0.0048\pm0.0023$ g/cm$^3$ at $P = 1.29$ kbar and $T = 214$ K. Thus the position of the LLCP is found to be at $P = 1.12\pm0.17$ kbar, $T = 215\pm1$ K. In previous studies \cite{ref34, ref35}, we estimated the critical pressure of the confined H$_2$O to be $1.6\pm0.3$ kbar by the dynamical properties of the system, including the dynamic crossover \cite{ref34} and the boson peak \cite{ref35}. Note that, the density measurement has a very good signal-to-noise ratio \cite{ref27, ref30}, thus its result is quite sensitive to the phase separation. However, the dynamic properties are not so sensitive. The changes of the dynamic properties, such as the disappearance of the dynamic crossover, only appear when the phase separation is significant enough. So it is not surprising that the critical pressure estimated by dynamic properties is higher than that obtained from the density measurement with elastic neutron scattering.

Furthermore, above the critical pressure, the maximum density difference increases as the pressure increases ($0.0048\pm0.0023$ g/cm$^3$ at $\sim$1.3 kbar; $0.010\pm0.003$ g/cm$^3$ at $\sim$1.6 kbar; $0.016\pm0.003$ g/cm$^3$ at $\sim$2.5 kbar), which agrees with an idea that in the vicinity of the critical point, the phase separation becomes more significant as the distance from the critical point increases along the liquid-liquid transition line.

We also tried other waiting times from 25 to 50 min for the density measurements at $\sim$1.6 kbar. The result shows that the value of the average density of the confined D$_2$O is effectively constant for different waiting times used here. This observation suggests that after waiting for 25 min, the sample temperature becomes stable and no evident transition happens up to 50 min.

From Fig. 9 one can find that, below about 1000 bar, no effective hysteresis is observed. This result is different from the result obtained from the continuous temperature scan \cite{ref27}. Such difference could be due to the temperature lag between the warming and cooling scans in the experiments with continuous temperature scan. In principle, the influence of the temperature lag on the density measurement has a positive correlation with the isobaric heat capacity of the confined water ($C_P$). Therefore, the hysteresis at low pressures may indicate the maximum of $C_P$. This conjecture can be justified as follows. According to relevant thermodynamic studies \cite{ref36, ref37}, at ambient pressure, the peak position of $C_P$ of the D$_2$O confined in MCM-41 with the pore diameter of 17\AA~is 240 K, which is very close to the temperature at which the maximum density difference takes place at ambient pressure in the result from the continuous temperature scan, 243 K (see Fig. 9 (b4)). The small difference between these two temperatures may be due to the difference of the pore diameter (according to Ref. \cite{ref36}, a $2$-\AA~increment in pore diameter can decrease the temperature of the peak of $C_P$ by several kelvins). Keep this idea in mind, one can then estimate the Widom line of the liquid-liquid transition, which is defined as the locus of the $C_P$ maxima in the corresponding one-phase region \cite{ref11}, with the positions of the maximum hysteresis observed in the continuous temperature scans at pressures lower than the critical pressure. Note that, in many other literatures, the Widom line is defined as the locus of the maximum correlation length \cite{ref10, ref38}. This definition can avoid the confusion introduced by the existence of multiple local maxima in the heat capacity of water \cite{ref39}. However, in this study, we still employ the former definition, since the heat capacity of the confined water is available and thus it is easy to compare our result to the result of thermodynamic measurement. In addition, as approaching the critical point, the maximum of heat capacity and the maximum of correlation length emerge \cite{ref10, ref38}.

Considering all the above discussions, we plot the phase diagram of the liquid-liquid transition of the confined heavy water in Fig. 10. The black solid squares denote the positions of the maximum density differences obtained by the continuous temperature scans at pressures higher than the critical pressure \cite{ref27, ref28}. These hysteresis phenomena cannot be completely eliminated by the discrete temperature scan protocol and denote the liquid-liquid transition of the confined water. By connecting these black solid squares with a smooth curve, and noting that the hysteresis disappears at pressures higher than 3500 bar in the temperature range from 140 to 300 K \cite{ref28}, we obtain the liquid-liquid transition line. The red open squares denote the positions of the maximum density differences obtained by the continuous temperature scans at pressures lower than the critical pressure \cite{ref27}. These hysteresis phenomena can be eliminated by the discrete temperature scan protocol and denote the positions of the $C_P$ maximum, \emph{i.e.} the Widom line. The liquid-liquid transition line and the Widom line intersect at $1.12\pm0.17$ kbar and $215\pm1$ K. This point could be the LLCP according to the LLCP scenario.

\begin{figure}
	\centering
\includegraphics{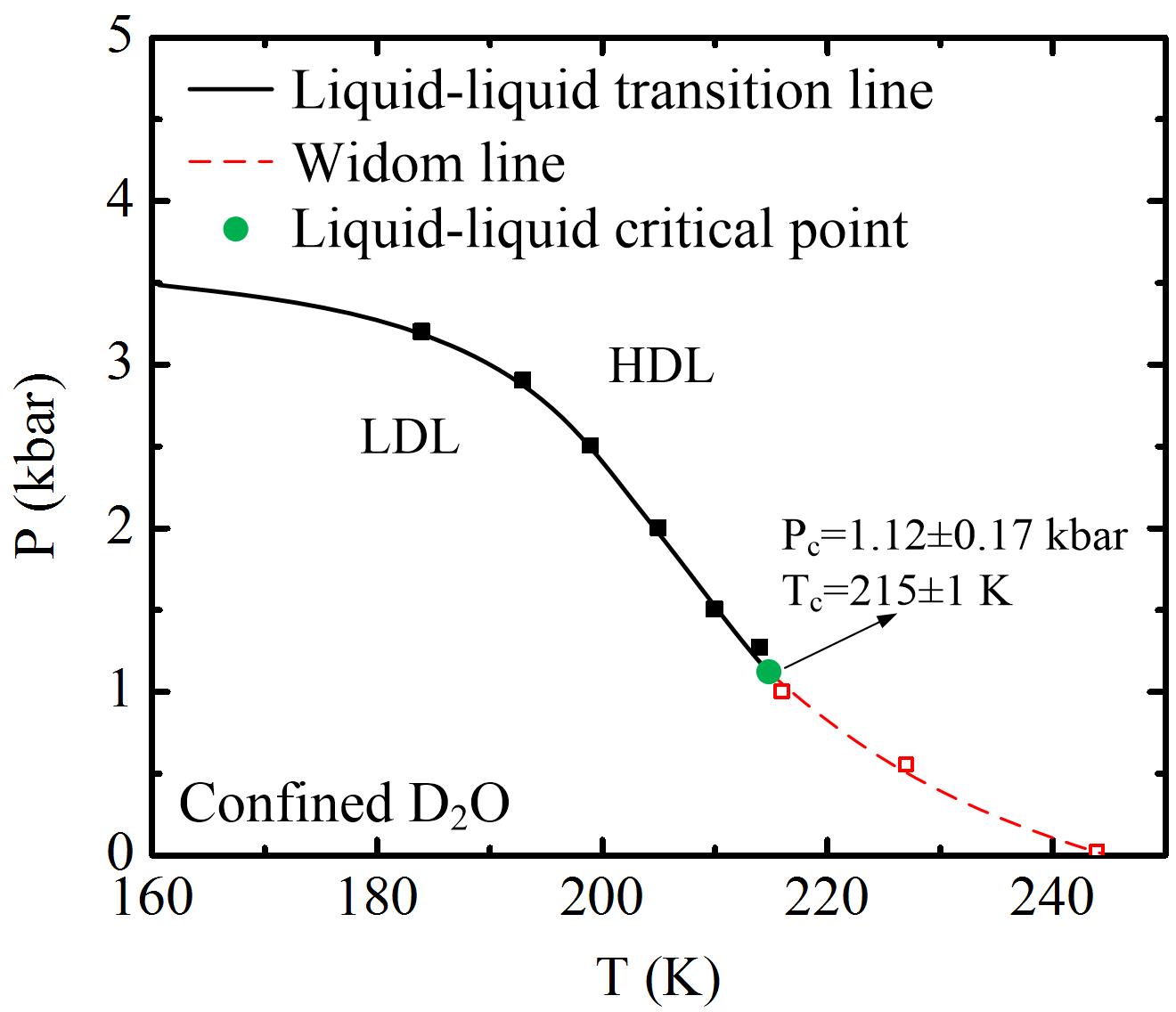}
\caption{Phase diagram of the liquid-liquid transition of the confined heavy water. The black solid squares and the red open squares denote the positions of the maximum density differences obtained by the continuous temperature scans at pressures higher than the critical pressure and lower than the critical pressure, respectively \cite{ref27, ref28}. The formers are due to the phase transition in the confined water, and represent the liquid-liquid transition line (denoted by a black solid line), while the latters are due to temperature lags, and represent the Widom line (denoted by a red dashed line). These two lines intersect at the LLCP, whose approximate position is denoted by a green circle.}
\end{figure}

It is believed that water undergoes glass transition at low temperatures \cite{ref40, ref41}. The transition temperature $T_g$ is conjectured to be at 136 \cite{ref42, ref43, ref44} or 165 K \cite{ref45} for bulk water, and 165 K for the water confined in MCM-41 at ambient pressure \cite{ref36}. All of these temperatures are much lower than the temperatures at which the hysteresis phenomena take place. Thus the hysteresis should not be directly induced by the possible glass transition in the confined water. Another concern is that due to the possible existence of the glass transition, below the conjectured $T_g$ the confined water may be in a glassy state, rather than an (metastable) equilibrium state, and the density measurement may be affected. In order to clarify this point, we perform a warming scan on density at 2 kbar by the following steps: first cool the system to 170 K at ambient pressure, then pressurize the system to 2 kbar and start the warming scan. In this route, the system temperature keeps on higher than the conjectured $T_g$ of the confined water and the system should be always away from a glassy state. This experimental route gives an effectively same density profile as compared to the one obtained by the warming scan starting from 140 K. Therefore we conclude that the hysteresis observed in this study is not affected by the possible glass transition in the confined water.

In order to examine the obtained phase diagram and to get a general idea on how the density of the confined water behaves as a function of $T$ and $P$, we perform isobaric density measurements on the confined D$_2$O at 5 pressures: 0.1 k, 1 k, 2.5 k, 4 k and 5 kbar. The data at 2.5 kbar are measured with warming scan. The results are shown in Fig. 11. According to the phase diagram shown in Fig. 10, below $\sim$190 K, the former 3 pressures are in the LDL phase, while the last 2 pressures are in the HDL phase. Figure 11 clearly shows that below 190 K, there is an evident density gap of $\sim$0.04 g/cm$^3$ between the density profiles at 0.1 k, 1 k, 2.5 kbar and the density profile at 4 kbar. This gap shows the phase separation between LDL and HDL. In this temperature range, the three density curves representing LDL phase are close to each other, which shows that the isothermal compressibility ($\chi_T$) of the LDL phase is small. At 170 K, the density only changes by $\sim$0.004 g/cm$^3$ as pressure increases from 100 bar to 2.5 kbar. In contrast, in HDL phase, the density changes by $\sim$0.02 g/cm$^3$ as pressure increases from 4 k to 5 kbar, which suggests a much larger $\chi_T$. Notice that, such a pressure dependence of $\chi_T$ is significantly different from that of a simple liquid. The local structure of the simple liquid is dominated by the excluded effect of the short-range repulsive interaction and has a relatively tight packing \cite{ref46}. Therefore, for a simple liquid, it is more difficult to compress it at higher pressures than at lower pressures.

The huge difference of $\chi_T$ in LDL and HDL, and the counterintuitive pressure dependence of $\chi_T$ of the confined water could be due to the different local structures of LDL and HDL. The LDL has a tetrahedral hydrogen-bond structure extending to the second coordination shell. While for the HDL, the second coordination shell collapses \cite{ref47}. These features make the local structure of the LDL more rigid and open than that of the HDL. Such sharp distinction on $\chi_T$ between HDL and LDL fades out as entering the one-phase region, which suggests that the LDL and HDL phases mix in this region.

\begin{figure}
	\centering
\includegraphics{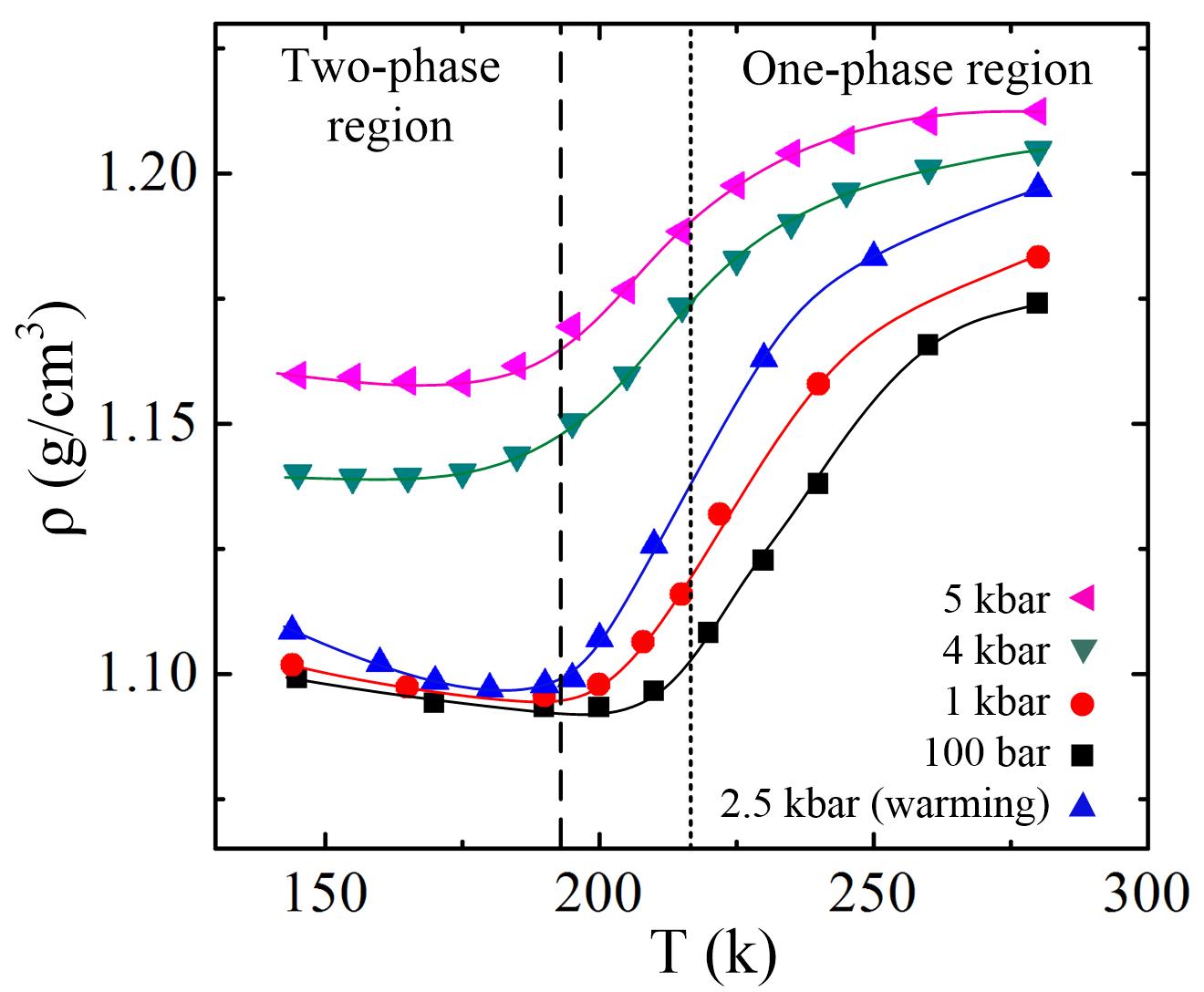}
\caption{The average density of the confined D$_2$O as a function of $T$ at $P$ = 0.1 k (black squares), 1 k (red circles), 2.5 k (heating scan, blue up triangles), 4 k (green down triangles) and 5 k (magenta left triangles) bar. The left-hand-side region of the dashed vertical line is the two-phase region with its phase separation between 3 and 4 kbar. The right-hand side region of the dotted vertical line is the one-phase region. From Z. Wang, \emph{et al.}, J. Phys. Chem. Lett. \textbf{6}, 2009 (2015).}
\end{figure}

Besides the LLCP scenario, the so-called singularity-free (SF) scenario \cite{ref48, ref49, ref50} also provides a phase diagram that is qualitatively similar to the one shown in Fig. 10. The SF scenario suggests that no singularity at the end point of the liquid-liquid transition, which differs from the LLCP scenario in which the liquid-liquid transition terminates with a critical point. To directly distinguish between these two scenarios one may want to study the critical behavior of this end point. Nevertheless, the quasi-one-dimensional geometry in MCM-41 can strongly suppress any critical behavior \cite{ref25}. Thus to measure the critical behaviors near the end point of the liquid-liquid transition is almost impossible. In fact, as the pressure approaches the critical pressure, the absolute value of the isobaric thermal expansion coefficient $|\alpha_P|$ of the confined D$_2$O exhibits no critical phenomenon \cite{ref51}. Kumar \emph{et al}. suggest another method to distinguish between these two scenarios: in the LLCP scenario the maximum of $C_P$ increases with the increase of pressure, while in the SF scenario the maximum of $C_P$ does not \cite{ref38}. In Ref. \cite{ref27}, below the critical pressure, the maximum density difference increases from 0.010 g/cm$^3$ at 1 bar to 0.017 g/cm$^3$ at $\sim$1 kbar. Considering that the $|\alpha_P|$ increases only by 2.7\% as $P$ increases from 1 bar to $\sim$1 kbar \cite{ref51}, we conjecture that such a big increase on maximum density difference as $P$ increases from 1 bar to $\sim$1 kbar is mainly due to the enhancement of the temperature lag, which indicates a larger $C_P$. Following this logic, we suggest that the LLCP scenario provides a better explanation. It is worth mention that, for bulk water, recent experimental and theoretical studies support the LLCP scenario rather than the SF scenario \cite{ref52, ref53, ref54}.

We compare the phase diagram of the confined heavy water \cite{ref30} and the conjectured phase diagram of the bulk heavy water \cite{ref55}, as shown in Fig. 12. The major difference is that the pressure of the liquid-liquid transition line of the confined water is higher than that of the bulk water by about 1 kbar. The reason might be attributed to the capillary effect due to the confinement in pores of cylindrical geometry. We use gas to pressurize the confined water system. For the fluid confined in a hydrophilic tube, the liquid-vapor surface forms a meniscus, which will lead to a pressure difference across this surface. Therefore, the pressure inside the nanopores and the pressure of the pressurizing gas can have a large difference. Here we use the Young-Laplace equation \cite{ref56} to give a rough estimation of the pressure difference for our case: when the tube is sufficiently narrow, the pressure difference can be expressed as $\Delta P = 2\gamma\cos\theta/R$. Where $\gamma$ is the surface tension of the fluid, $\theta$ is the contact angle and $R$ is the radius of the tube. With the values of $\gamma$ and $\theta$ at room temperature, the pressure difference for the nanopores of MCM-41 is estimated to be in the order of 1000 bar, which is roughly consistent with the pressure difference between the results of the bulk water and the confined water. Note that, Young-Laplace equation cannot describe the liquid confined in nanopores accurately. Further studies on the pressure effect of the cylindrical nanopore are necessary. These studies may provide a link between the bulk water and the confined water.

\begin{figure}
	\centering
	\includegraphics{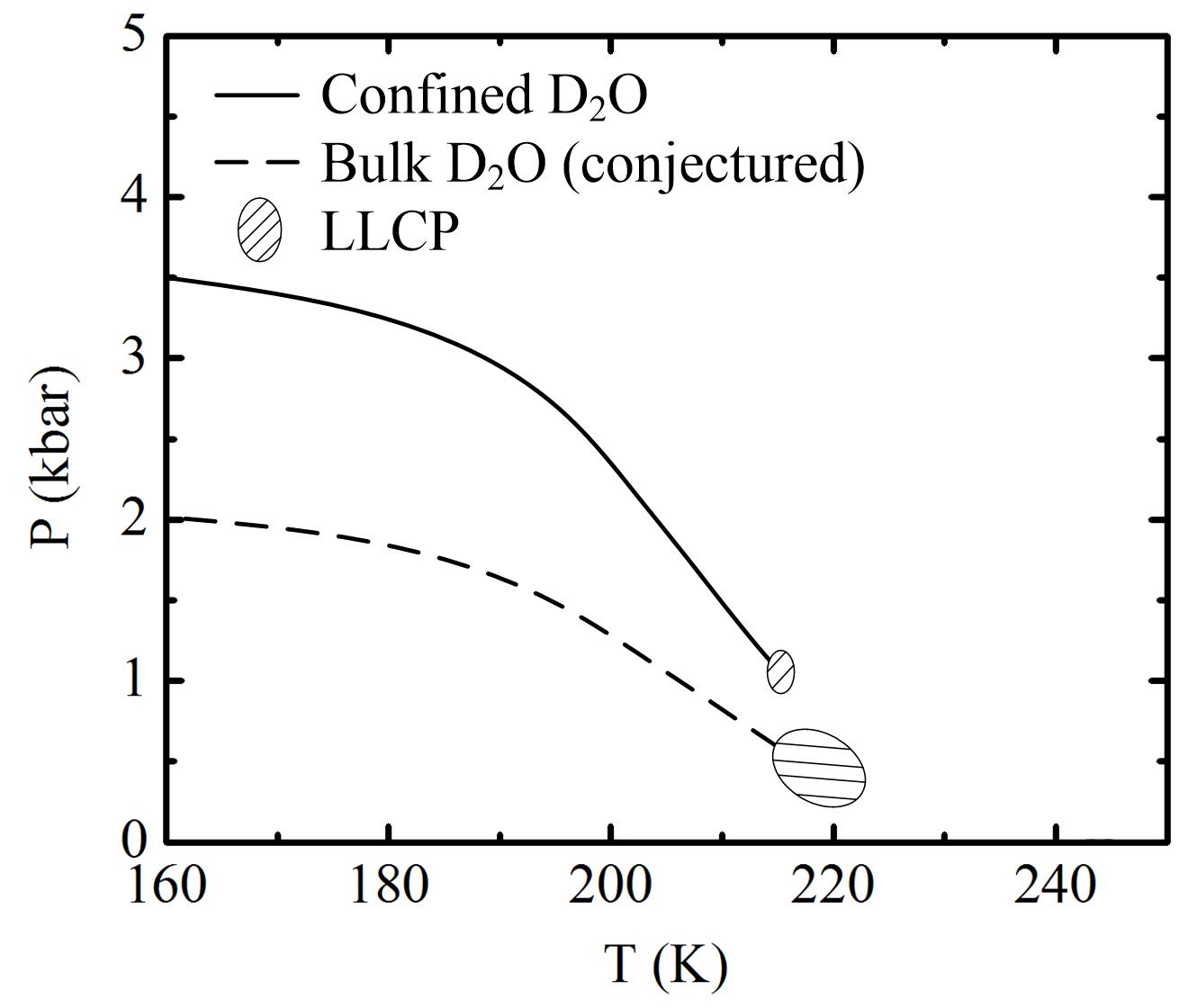}
	\caption{Comparison between the phase diagram of the confined D$_2$O (solid line) \cite{ref30} and the conjectured phase diagram of the bulk D$_2$O (dashed line) \cite{ref55}.}
\end{figure}

\subsection{Partially-Hydrated Sample}

In 2008, Liu \emph{et al}. measured the average density of the confined heavy water in a partially-hydrated sample as a function of temperature at ambient pressure \cite{ref19}. The hydration level of this sample is 85\% of the full hydration level. The result is shown in Fig. 13. It can be found that, for this sample, the density minimum obscures and the maximum of the absolute value of the isobaric thermal expansion coefficient ($|\alpha_P|$) decreases as compared to the fully-hydrated sample. The result suggests that the confined water in the partially-hydrated sample is not as ``anomalous'' as the one in the fully-hydrated sample. Therefore, it is necessary to examine if a reduction of hydration level can mitigate the phase transition at high pressures.

\begin{figure}
	\centering
\includegraphics{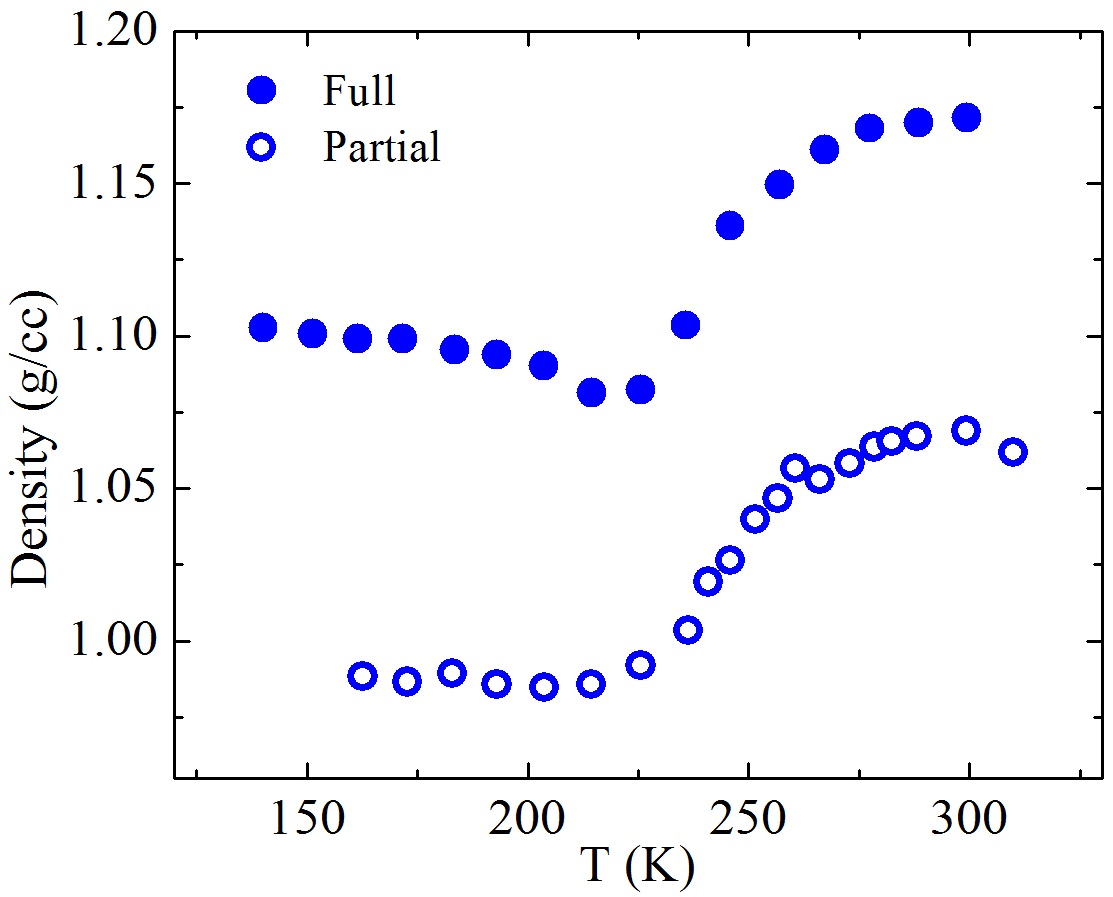}
\caption{Average density of the confined D$_2$O as a function of temperature at ambient pressure. The solid circles and open circles denote the results of the fully-hydrated sample and the partially-hydrated sample, respectively \cite{ref19}. The D$_2$O density in the partially-hydrated sample is lower than that of the fully-hydrated sample because of the existence of a partially empty central core. From D. Liu, \emph{et al.}, J. Phys. Chem. B \textbf{112}, 4309 (2008).}
\end{figure}

Recently, we studied the average density of the confined heavy water of an 80\% partially-hydrated sample at $\sim$1.6 kbar with warming and cooling scans \cite{ref30}. The result is shown in Fig. 14. Strikingly, the density hysteresis completely disappears in this sample. The disappearance of the density hysteresis in the partially-hydrated sample is also observed at 1 k and 2.5 kbar. Notice that, both experimental and computer simulation studies show that the confined water has a layer structure \cite{ref13, ref57, ref58}. According to Gallo \emph{et al}. \cite{ref13}, the water confined in MCM-41 can be divided into two dynamically distinct parts in radial direction: bound water and free water. The bound water is a 3-\AA-thick shell layer that coats to the hydrophilic surface of the silica cavity, while the free water is the water in the center part of the cavity. Since the water forms the shell layer first \cite{ref59}, the 20\% lowering of $h$ is mainly due to the reduction of the free water. Thus, in this partially-hydrated sample, the amount of free water decreases by about 50\% compared to its fully-hydrated counterpart. The disappearance of the density hysteresis in the partially-hydrated sample strongly suggests that (1) the free water, not the bound water, undergoes a liquid-liquid transition, and (2) a well-developed hydrogen-bond network in free water is the necessary condition for water confined in MCM-41 to exhibit liquid-liquid transition.

\begin{figure}
	\centering
\includegraphics{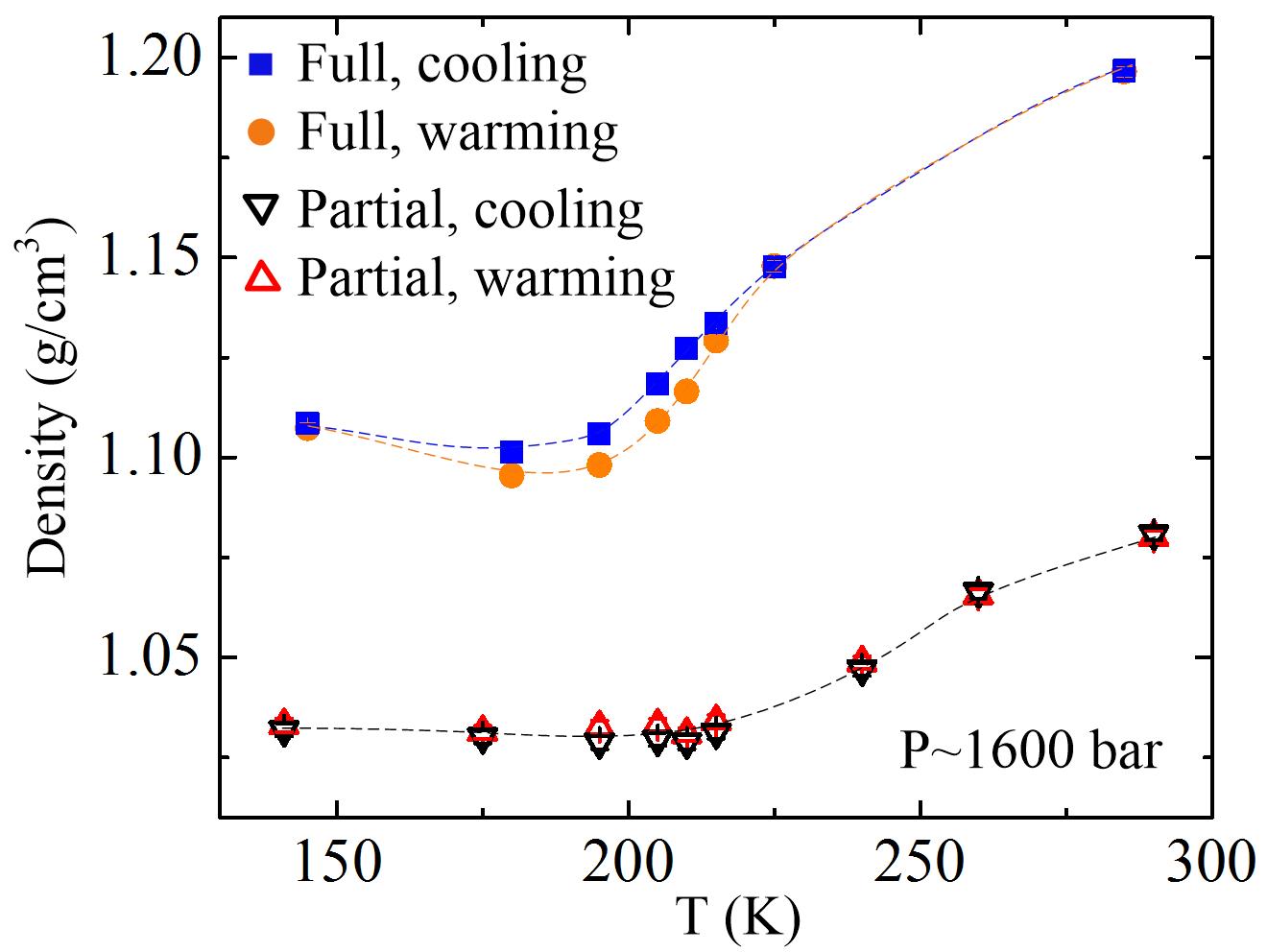}
\caption{Density profiles of confined D$_2$O with warming (red open up triangles) and cooling (black open down triangles) scans for a partially-hydrated sample at $P \sim$ 1600 bar. It is seen that no hysteresis is found for this sample. We also plot the density profiles of confined D$_2$O with warming (orange circles) and cooling (blue squares) scans for the fully-hydrated sample at $P \sim$ 1600 bar for comparison. From Z. Wang, \emph{et al.}, J. Phys. Chem. Lett. \textbf{6}, 2009 (2015).}
\end{figure}

Since the free water is the part which undergoes a liquid-liquid transition, it is important to study the density behavior of the free water as a function of pressure and temperature. We can make a rough estimation by assuming that (1) the thickness of the bound water keeps as 3 \AA, and (2) the temperature dependence of the density of the bound water is similar to that of the confined water in the partially-hydrated sample. The result is shown in Fig. 15. 

\begin{figure}
	\centering
\includegraphics{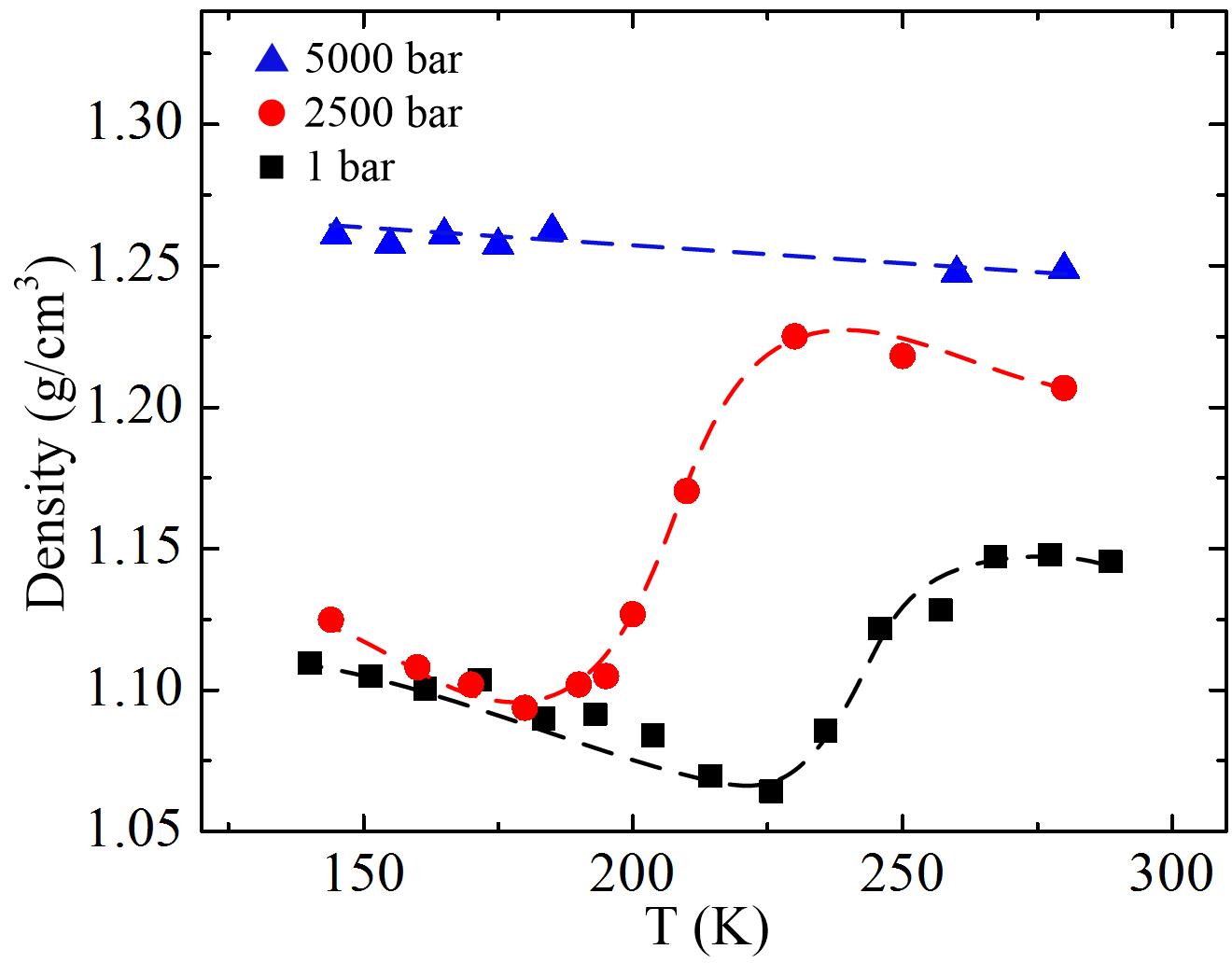}
\caption{Semi-quantitative estimations of the density of the free water as a function of temperature at different pressures. The black squares, red circles and blue triangles denote the results at 1, 2500 and 5000 bar, respectively. The dashed lines are to guide eyes.}
\end{figure}

Figure 15 suggests that at pressures higher than about 5 kbar, the density of the free water as a function of temperature behaves like a normal liquid, namely, the density of the free water increases as the temperature decreases. This is not unexpected, because at such high pressures, the confined water is dominated by HDL phase, which is more ``normal'' than the LDL phase. The estimation on the density behavior of the free water shown in Fig. 15 is qualitatively similar to the density behavior of the bulk water \cite{ref53, ref54}. For bulk water, when the pressure is higher than about 2 kbar at where the water is supposed to be dominated by HDL, the density as a function of temperature does not exhibit maximum and minimum. It increases monotonically as the temperature decreases. Notice that, the pressure difference between the confined water case (5 kbar) and the bulk water case (2 kbar) should be due to the pressure effect of the nanoscale confinement, which has been discussed in the above section.

We emphasize that, the result in Fig. 15 is only a semi-quantitative estimation based on the above-mentioned two assumptions. In principle, the density of the free water can be obtained by measuring the fully-hydrated sample and partially-hydrated sample, and then calculating their difference. However, to get an accurate value of the density of the free water from elastic neutron scattering experiment is practically difficult, mainly because (1) the thickness of the bound water and its temperature and pressure dependences are not available, and (2) the effect of the interaction between the bound water and free water is unclear. Further investigations in simulation could aid this point.

\section{Concluding Remarks}
This work summarizes our recent work on the detection of the liquid-liquid transition in the confined water with elastic neutron scattering [18, 19, 27, 28, 30]. The observations of the density minimum and the density hysteresis in the deeply-cooled region are remarkable, since these phenomena are consequences of the hypothetical liquid-liquid transition. The absence of the hysteresis in the partially-hydrated sample provides further insight into the detail of this likely first-order transition. It suggests that the bound water, whose properties are strongly influenced by the surface chemistry of the confining material, does not exhibit the transition. In contrast, the free water, which is less influenced by the confining material and has a stronger hydrogen-bond tetrahedral network than the bound water, undergoes the transition at high pressures. The phase diagram of the confined heavy water was established by density hysteresis measurements with two kinds of temperature changing protocol. In Fig. 16, we present it with gradual color change to give a visual picture of this transition.

\begin{figure}
	\centering
\includegraphics{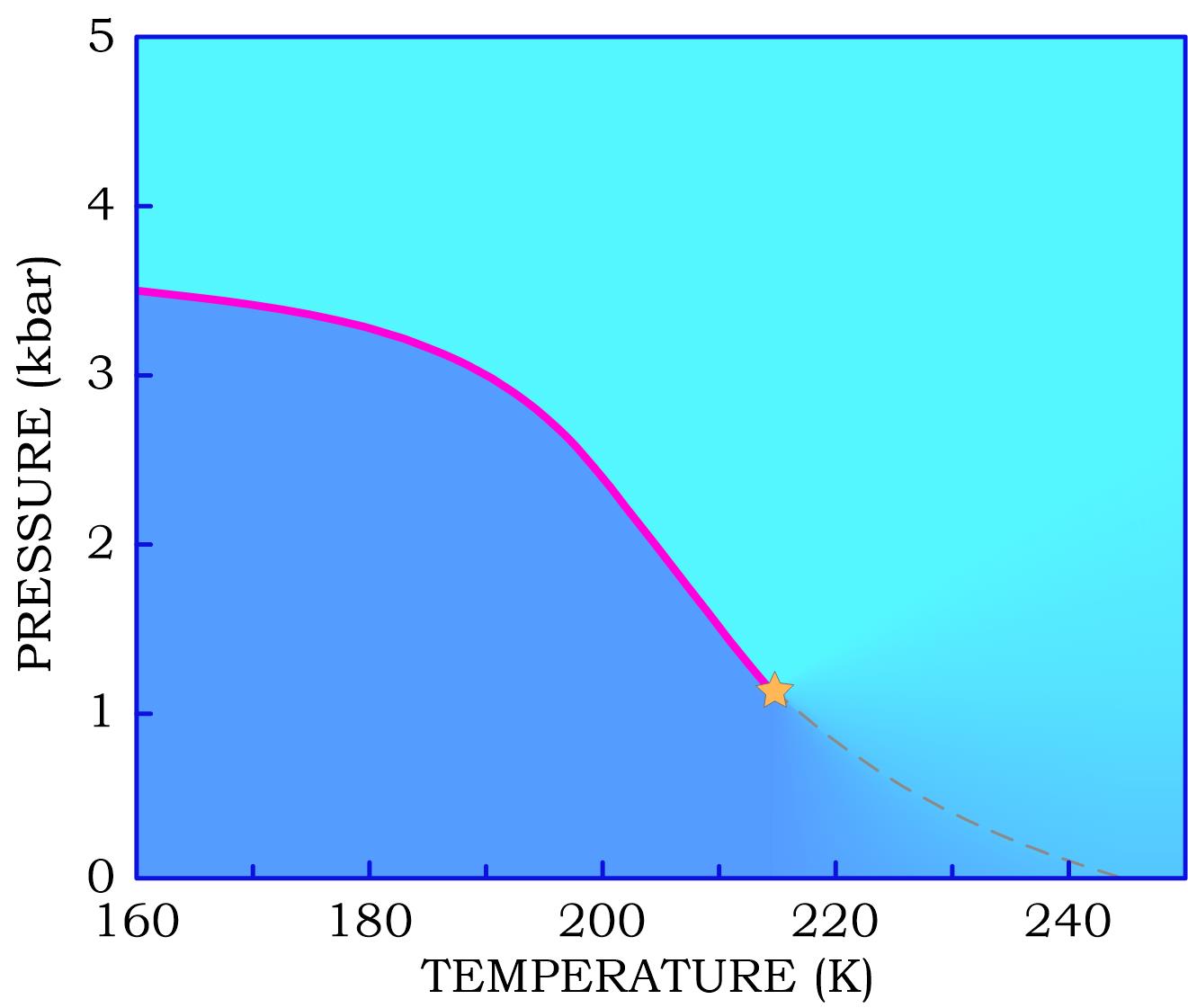}
\caption{Phase diagram of the liquid-liquid transition in the confined heavy water. The cyan and blue colors represent the HDL-dominant and LDL-dominant regions, respectively. The solid line denotes the liquid-liquid transition line. The dashed line denotes the Widom line. The pentacle denotes the LLCP.}
\end{figure}

The study of the phase behavior of the confined water was originally motivated by the anomalous properties of the bulk water and relevant theoretical and computer simulation studies. In fact, the confined water itself is interesting and important as well. Similar to bulk water, confined water exhibits thermodynamic anomalies in density \cite{ref15, ref18, ref21, ref60}, isobaric expansion coefficient \cite{ref15, ref18}, isobaric heat capacity \cite{ref36} and isothermal compressibility \cite{ref30} at low temperatures. Many of its transport properties also behave differently from normal liquids. For instance, the diffusion coefficient and viscosity of the confined water break the Stokes-Einstein law at about 220 K \cite{ref61}. Moreover, the characteristic relaxation time of the confined water exhibits a large and unusual decrease as pressure increases from 1 bar to 4 kbar at temperatures lower than 230 K \cite{ref62}. We argue that, all these anomalous phenomena, together with the density hysteresis at high pressures, can be understood by accepting the existence of the liquid-liquid transition in the deeply-cooled region of the confined water.

Besides confined water, people have prepared other systems to enter the ``no man's land'' and to detect the existence of the liquid-liquid transition. Different aqueous solutions have been studied by kinetic measurement \cite{ref23, ref24} and thermodynamic measurement \cite{ref63}. It is found that there are two liquid phases exist in such systems that differ in density and structure. These two phases correspond to the LDL and HDL. For the detection of the liquid-liquid transition in bulk pure water, a breakthrough was made recently. In 2014, Nilsson and his collaborators reported data on the structure of liquid water well below the homogeneous nucleation temperature $T_\mathrm{H}$ \cite{ref64}. They prepared a jet of water droplets with the length scale of 10$^{-5}$ m and measured their structures with ultra-fast X-ray diffraction technique. The result shows that at the temperature of 227 K, the local structure of liquid water is drastically changed from the local structure of water at ambient conditions. In this case, water is almost a tetrahedrally structured liquid, which is the signature of the LDL phase.

The concepts of HDL and LDL, after being introduced by Eugene Stanley and his collaborators twenty-three years ago, attracted a great deal of attention. The idea that two structurally different phases can exist in a one-component liquid is profound. In last two decades, researchers from different disciplines, using computer simulations and experimental methods, attack the ``no man's land'' and try to find clues for the liquid-liquid transition. Though a unified opinion is still lacking, all of these efforts represent steps towards an ultimate understanding of the unique behavior of liquid water. 

\acknowledgments
The research at MIT was supported by DOE grant DE-FG02-90ER45429. We thank Dr. Leland Harriger, Dr. Juscelino B. Le\~ao and Peisi Le for their help in this project.

\bibliography{ref}{}

\begin{thebibliography}{10}
\bibitem{ref17}
Notice that the pore size determined by BJH method may underestimate a more
  geometric definition of the pore diameter.
\bibitem{ref29}
For the measurements at 3.3 and 4 kbar, we first performed the continuous
  cooling scan from 300 K to 140 K. After two hours of waiting, we then
  performed the continuous warming scan from 140 K to 300 K.
\end{thebibliography}
\bibliographystyle{varenna}
\end{document}